\documentclass[twocolumn, conference]{IEEEtran}
\usepackage{epsfig}
\usepackage{verbatim}
\usepackage{ulem}
\usepackage{amsmath}
\usepackage{textcomp}
\usepackage{footnote}
\usepackage{multicol}
\usepackage{graphicx}

\ifx\pdfoutput\undefined
\usepackage[dvips]{color}
\else
\usepackage{color}
\fi

\renewcommand{\baselinestretch}{1.0}
\usepackage{hyperref}
\hypersetup{%
  unicode   =false,    
  pdftoolbar=false,    
  pdfmenubar=true      
}

\begin{document}
\pagestyle{empty}

\setlength\parindent{0pt}

\makeatletter
\def\ps@headings{%
\def\@oddhead{\mbox{}\scriptsize\rightmark \hfil \thepage}%
\def\@evenhead{\scriptsize\thepage \hfil \leftmark\mbox{}}%
\def\@oddfoot{}%
\def\@evenfoot{}}
\makeatother

\title{An Efficient Transition Algorithm For Seamless Drone Multicasting}
\thispagestyle{empty}
\author{Wanqing Tu \\
School of Computer Science, The University of Auckland, New Zealand. \\
Email: w.tu@auckland.ac.nz} \maketitle
\pagestyle{empty}\thispagestyle{empty}
\begin{abstract}
Many drone-related applications (e.g., drone-aided video capture, drone traffic and safety management) require group communications between drones to efficiently disseminate data or reliably deliver critical information, making use of the line-of-sight coverage of drones to realise services that ground devices may not be capable of. This paper studies high-performance yet resource-efficient mobile drone multicasting via trajectory adjustment. We first analyse the trajectory adjustment condition to determine whether a straight-line trajectory is fully covered by the multicast or not, by conducting simple computation tasks and with controlled overhead traffic. We then propose the trajectory adjustment scheme to provide a new trajectory with controlled travel distances. The ETTA algorithm is finally presented to apply the trajectory adjustment condition and scheme to a drone transiting between forwarders whose coverage do not overlap. The algorithm relies on multicasting forwarders, instead of additional transition forwarders, to fully cover the adjusted trajectory, helping to control interference and network traffic load. Our NS2 simulation results demonstrate that ETTA, as compared to other mobile multicasts, can achieve guaranteed performance for drone receivers in a multicast with heavier traffic loads.
\end{abstract}
\section{Introduction}
To enable information to be delivered to a group of drones is essential for many drone-related applications and services, including drone safety monitoring, drone traffic management, air rescue assistance, surveillance operations, aerial video capture for professional or entertainment purposes, etc. Multicasting between drones is an efficient communication method for such group applications, not only because it may deliver data to multiple receivers with much less use of transmission resources but also because it can deliver critical information (e.g., safety alarms, traffic schedules) reliably via multiple parallel paths. Moreover, drone multicasting is helpful in relaying information for ground multicasting in which members are not within line of sight of each other. In these group applications, drones often change their locations in order to capture information from different angles, adjust line of sight ranges between sky and ground, etc. This paper studies mobile drone transitions in aerial multicasting, supporting the development of relevant drone applications.

Mobile drone multicasting inherits the challenges faced by mobile multicasting on the ground: node transitions cause interrupted connections or increased interference. Conventional ground solutions enhance tree-based or mesh-based multicasting protocols [1-2] to avoid interruption or interference. They however introduce considerable traffic overheads or complex maintenance for networks and their devices, making them unsuitable for drones that often have energy constraints and computational limitations. More popular studies develop geographic multicasting [3-6]. 
In general, geographic multicasting arranges group members into different physical zones based on their locations. Multicasting trees are established to connect different zones and broadcast is used within each zone. Zone leaders manage members' joining or leaving which decreases the requirement to adjust multicasting trees between zones, greatly reducing the incidence of interrupted mobile connections and the associated interference and overheads. However, broadcasting unnecessarily spreads data to nodes that do not belong to a group, wasting wireless bandwidth and device energy. Moreover, it is complicated to plan or form zones based on nodes' physical locations.

For drone-related multicasting, the literature mainly focuses on drone-to-earth applications (e.g., [7-8]) in which a drone disseminates data to a set of ground users/devices. Multicasting between drones is rarely studied. The resource limitations of drones and their wireless connections require handover operations to make light use of resources while guaranteeing seamless transitions. Therefore, drone transitions with low traffic overhead, controlled energy consumption, and simple computation tasks form our design target. Our transition takes advantage of the existence of multiple forwarders in a multicasting as well as an obstacle-free aerial communication environment to achieve the expected transition performance. In our system, straight-line trajectories are adopted with priority to transfer drones, incurring short travel distances and low traffic overheads and hence benefitting resource efficiency. However, straight-line trajectories may not always be seamless. We hence design a new algorithm - efficient transition via trajectory adjustment (ETTA) to guide drones' movement with controlled resource consumption. In detail, our contributions include the following results.
\begin{itemize}
\item Trajectory adjustment condition. We analyse the condition when a straight-line trajectory is not fully covered by the multicasting system if a drone transits between forwarders that have overlapping coverage. The implementation of such condition should be a fast yet resource-efficient process as it mostly requires to calculate Euclidean distance based on the 3-dimensional Pythagorean theorem.
\item Efficient trajectory adjustment schemes. We design new trajectories to replace interrupted straight-line trajectories for transitions between forwarders with or without overlapping coverage. To balance the tradeoff between enabling a fast transition and controlling resource utilisation, these new trajectories are established to limit the extra travel distance exceeding that of the original straight-line trajectory when drones move between forwarders with overlapping coverage, or to employ a minimal number of forwarders from the multicasting structure that together can provide seamless coverage for drones when they move between forwarders without overlapping coverage.
\item The ETTA algorithm. It systematically combines the condition and schemes of trajectory adjustment, supporting efficient yet seamless drone transitions in aerial multicasting.
\end{itemize}

Finally, we use NS2 simulations to evaluate our ETTA. We observe the average multicast delays, the average multicast throughput, and the average mobile throughput in different multicasting networks. The results show that ETTA may admit $64\%$ more traffic load while guaranteeing the multicasting performance for both mobile and stable drone receivers.

\section{Related Studies}
Wireless multicast with static group members focuses on improving complex interference and limited wireless bandwidth. Early strategies avoid interference by utilising non-overlapping channels between nearby nodes (e.g., [9]) or by hopping nodes between different channels (e.g., [10]). Transmission scheduling is another well studied strategy that efficiently utilises channel resources to gain more transmission opportunities. Studies have scheduled transmission rates (e.g., [11]), flow transmissions (e.g., [12]), etc. to enable a channel to accommodate more multicasts or to extend multicasts' coverage. External resources, such as licensed RF bands (e.g., [13]) or wired network links (e.g., [14-15]), are also exploited for additional bandwidth.

For mobile multicasting, many studies concern the reliable performance received by mobile members. In [16], a tree multicast is designed that assigns an ID to each multicasting node. Flows are forwarded in order of IDs. Interrupted connectivity is repaired by referring to the sparseness between IDs. The core-assisted mesh protocol [1] builds a shared multicast mesh to maintain group connectivity when network routers move frequently. It reverses the shortest unicast paths to form multicasting paths on the shared mesh, supporting loop-free packet forwarding. Tree- or mesh-based mobile multicasting often requires complex operations to maintain connectivity, generating considerable overheads to bandwidth-limited mobile networks. Geographic multicasting (e.g., [3]) improves this drawback by dividing group members into different zones. These zones are connected via a multicasting. This multicasting structure does not change with nodes' mobility because their movements do not cause zone movement. Within each zone, data is delivered via greedy forwarding. The geographic multicasting protocol in [4] computes a Steiner tree to connect zones. It carries concurrent multicasts for a higher delivery ratio, resulting in scalable delay performance even when network sizes increase. The work in [17] designs a virtual-zone-based structure to manage group members. With the position information of multicasting nodes, it constructs a zone-based bidirectional tree. The protocol uses zone depth to optimise tree structures and integrates nodes' location information with group member management to enhance multicasting efficiency.

For drone-related multicasting, in [7], drone-to-earth multicast transmissions are developed by using filter bank multicarrier. The proposal designs filter bank multicarrier with offset quadrature amplitude modulation (offset-QAM) and hermite polynomial-based prototype filtering, helping to manage the tradeoff between performance and spectral efficiency. In [8], drone trajectories are designed theoretically to minimise mission completion time while ensuring each ground terminal to recover the file with a high probability. In general, while multicasting between drones is important to support many emerging, it is however rarely studied in literature.

\section{Efficient Transition Via Trajectory Adjustment (ETTA)}
In our drone multicasting system, we employ the link-controlled routing tree (LCRT) algorithm [14-15] (illustrated in Fig.~\ref{fig3} of Section III.B) to form interference-controlled paths between drones. This section studies how to seamlessly transit drones in a multicasting. We first explore drone transitions between overlapping forwarders, i.e., two forwarders with overlapping coverage. We then study transitions between non-overlapping forwarders, i.e., forwarders without overlapping coverage. On this basis, the ETTA algorithm is presented.

\subsection{Transitions Between Overlapping Forwarders}
\begin{figure}[h]
\begin{center}
\begin{tabular}{c}
\includegraphics[trim=40 420 40 130, clip,height=1.65in]{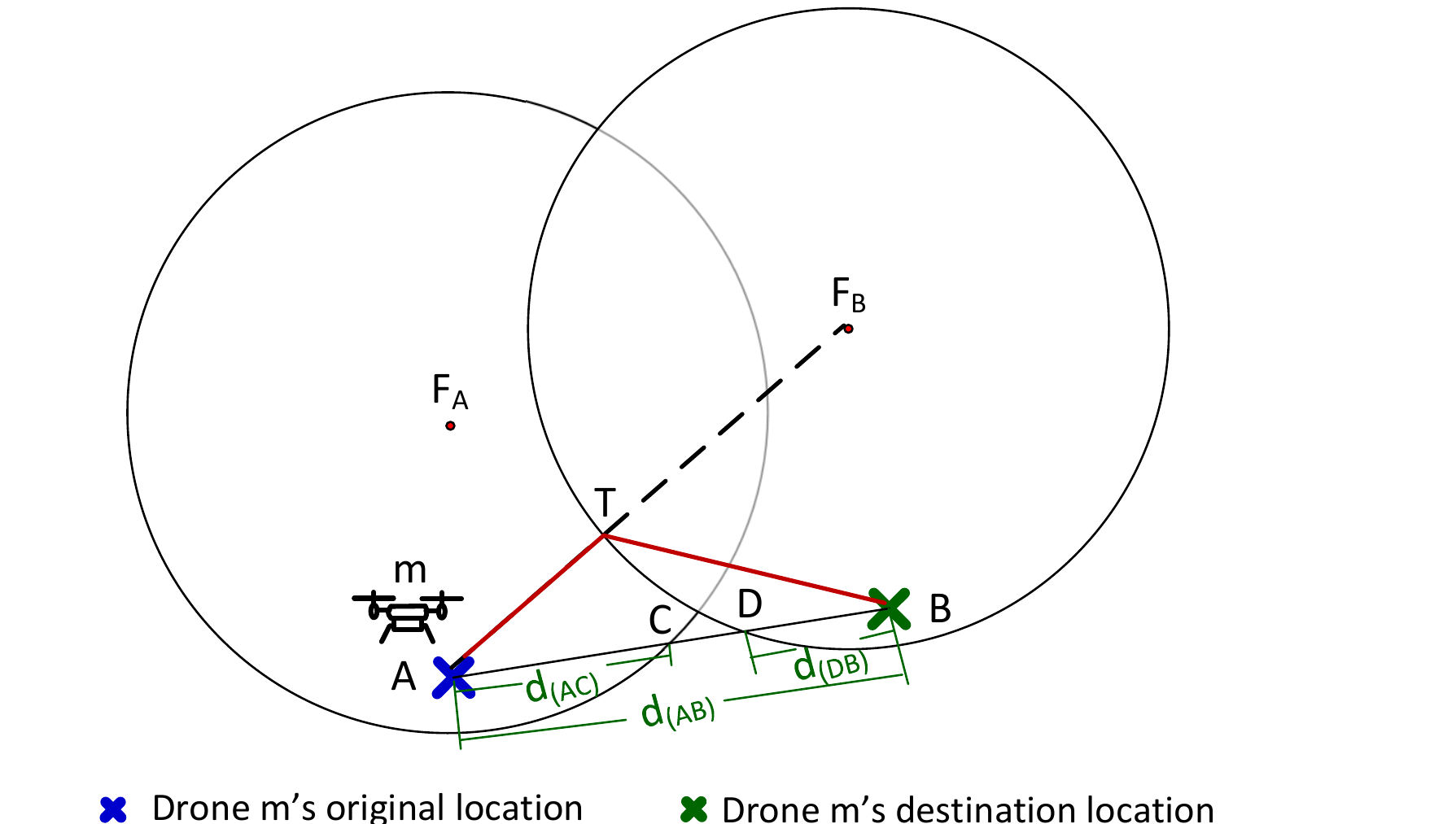}
\end{tabular}
\end{center}
\caption{Deriving the trajectory adjustment condition.}\label{tac}
\end{figure}
As mentioned, straight-line trajectories support fast drone transitions and use resources efficiently. However, they may not always be seamless. We use Fig.~\ref{tac} to illustrate how to determine whether a straight-line trajectory is seamless or not when the mobile drone $m$ transits between two overlapping forwarders. In our system, as drones communicate via omnidirectional antennas in sky space, we assume that a drone's transmission range is a sphere with the radius of $r$. In Fig.~\ref{tac}, if $m$ moves from $A$ to $B$, let $m$'s original and destination forwarders be $F_A$ and $F_B$, and the intersections of the straight-line trajectory with the coverage edges of $F_A$ and $F_B$ be $C$ and $D$ respectively. Denote the distances between $A$ and $B$, $A$ and $C$, and $B$ and $D$ as $d_{(AB)}$, $d_{(AC)}$, and $d_{(BD)}$ respectively. Theorem 1 gives the trajectory adjustment condition.

{\bf Theorem 1.} {\it For a drone $m$ moving between two overlapping forwarders (shown in Fig.~\ref{tac}), the straight-line trajectory from its origin $A$ to its destination $B$ is seamless if one of the following conditions meets: 1) $d_{(AB)}\leq{d_{(AC)}+d_{(DB)}}$, or 2) when $d_{(AB)}>{d_{(AC)}+d_{(DB)}}$, there exists a forwarder in the multicasting whose distances to $C$ and $D$ are both $\leq{r}$. \\
Otherwise, the straight-line trajectory needs to be adjusted.}

{\bf Proof.} We prove Theorem 1 by contradiction. When $d_{(AB)}\leq{d_{(AC)}+d_{(DB)}}$, suppose the straight line $A\rightarrow{B}$ is not seamless. Then, some part(s) of the straight-line trajectory is(are) not covered by $F_A$ and $F_B$. Let the length of the uncovered part(s) be $l>0$. We have
$
d_{(AC)}+l+d_{(DB)}=d_{(AB)}\Rightarrow{l=d_{(AB)}-d_{(AC)}-d_{(DB)}}.
$
Since $l>0$, we have
$
d_{(AB)}>{d_{(AC)}+d_{(DB)}}.
$
This contradicts $d_{(AB)}\leq{d_{(AC)}+d_{(DB)}}$. Therefore, when $d_{(AB)}\leq{d_{(AC)}+d_{(DB)}}$, the straight-line trajectory does not need to be adjusted.

When $d_{(AB)}>{d_{(AC)}+d_{(DB)}}$, suppose there exists a multicasting forwarder $f$ whose distances to $C$ and $D$ are both $\leq{r}$. If the straight-line trajectory is not seamless, there is at least a point between $C$ and $D$ whose Euclidean distance to $f$ is $>r$. This makes that $C\rightarrow{D}$ is not a straight line because the two ends $C$ and $D$ are both within the distance of $r$ to $f$, contradicting the fact that
$A\rightarrow{B}$ is a straight line. Q.E.D

The implementation of Theorem 1 requires knowledge of the Euclidean coordinates of $C$ and $D$, denoted as $(x_C, y_C, z_C)$ and $(x_D, y_D, z_D)$ respectively. 
As $C$ is on the edge of $F_A$'s transmission range. We have
\begin{equation}
(x_C-x_{F_A})^2+(y_C-y_{F_A})^2+(z_C-z_{F_A})^2=r^2.
\end{equation}
Also, $C$ is on the straight-line trajectory $A\rightarrow{B}$, i.e.,
\begin{equation}
\begin{cases}
x_C=x_A+t(x_B-x_A), \\
y_C=y_A+t(y_B-y_A), \\
z_C=z_A+t(z_B-z_A).
\end{cases}
\end{equation}
Inputting (2) into (1), we obtain $t$ by solving the quadratic equation for $t$. Typically two distinct values of $t$ will be obtained, defining two distinct points. The point closer to $F_B$ is $C$. Similarly, we obtain $D$'s coordinates. Then, by Theorem 1, if $A\rightarrow{B}$ is seamless, $m$ transits via this trajectory. Otherwise, a new trajectory is formed as below.

When proposing a new trajectory, we try to control travel distances and traffic overheads with computation of low complexity, allowing fast transitions with efficient use of resources (e.g., energy, bandwidth). The idea is to employ a location (denoted as $T$), within the overlap of transmission ranges of $F_A$ and $F_B$, to form a transition path $A\rightarrow{T}\rightarrow{B}$ inside the combined coverage of $F_A$ and $F_B$. Ideally, $T$ should minimise the extra travel distance exceeding that of the straight-line trajectory. Such a location is achievable by an existing algorithm (e.g., [19]) to seek a point on the surface of the overlapping area that has the shortest distance to the straight line $A\rightarrow{B}$. However, this potentially increases computation delays and its energy consumption. 
Therefore, as illustrated by $T$ in Fig.~\ref{tac}, we use the closest intersection between the line $A\rightarrow{F_B}$ and the edge of 
$F_B$'s coverage. 
$T$'s coordinates can be represented as below, by the line function between $A$ and $F_B$,
\begin{equation}\label{lineequation}
\begin{cases}
x_T = t(x_{F_B}-x_A)+x_A, \\
y_T = t(y_{F_B}-y_A)+y_A, \\
z_T = t(z_{F_B}-z_A)+z_A. \\
\end{cases}
\end{equation}
As $T$ is on the edge of $F_B$'s coverage, we have $(x_T-x_{F_B})^2+(y_T-y_{F_B})^2+(z_T-z_{F_B})^2=r^2$. Combining this equation with (3), we can derive $t$ and hence $T$'s coordinates. The new trajectory $A\rightarrow{T}\rightarrow{B}$ is shown by the red lines in Fig.~\ref{tac}.

\subsection{Transitions Between Non-Overlapping Forwarders}
\begin{figure}[h]
\begin{center}
\begin{tabular}{c}
\includegraphics[trim=10 460 5 2, clip,height=1.9in]{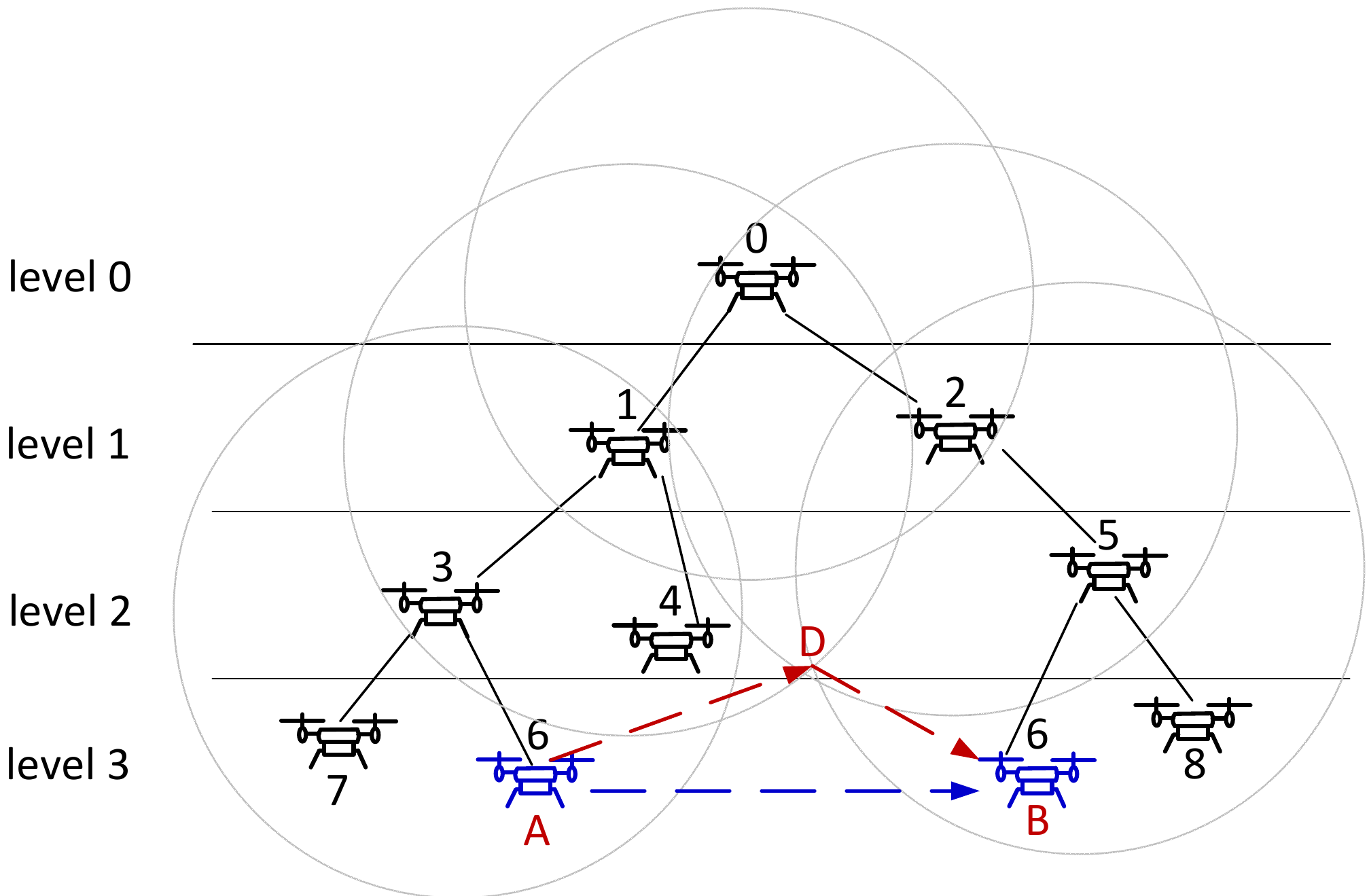}
\end{tabular}
\end{center}
\caption{An example of forming the LCRT tree and the ETTA trajectory.}\label{fig3}
\end{figure}
Recall that, in our system, a multi-hop multicasting tree is established by the LCRT [14-15] algorithm to connect drones. As shown in Fig.~\ref{fig3}, 
drones are assigned to different levels based on their shortest hop distances to the source (i.e., drone 0 in Fig.~\ref{fig3}). 
Then, from the second highest level (level 2 in Fig.~\ref{fig3}) to the second lowest level (level 1 in Fig.~\ref{fig3}), at each level, drones covering more forwarders/receivers at the immediately higher level are selected as forwarders with priority, until all forwarders/receivers at the immediately higher level have found their forwarders.

\begin{figure}[h]
\begin{center}
\begin{tabular}{c}
\includegraphics[trim=100 530 150 132, clip,height=1.1in]{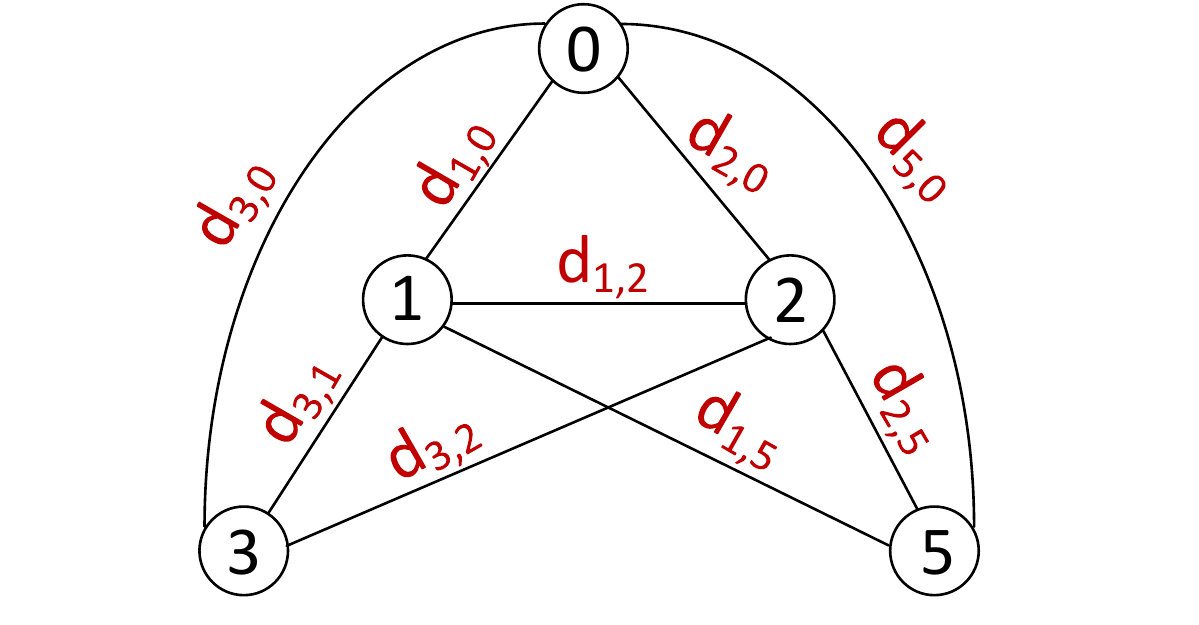}
\end{tabular}
\end{center}
\caption{The overlapping graph of the LCRT multicasting tree in Fig.~\ref{fig3}.}\label{graph}
\end{figure}

When transiting between non-overlapping forwarders, in order to form a seamless trajectory with controlled travel distance to replace an interrupted straight-line trajectory (e.g., the blue dotted line in Fig.~\ref{fig3}), our idea is to select a minimal number of LCRT forwarders that overlap one by one to provide coverage along the transition path. In detail, $m$ generates an overlapping graph to represent how LCRT forwarders' coverage overlaps. LCRT forwarders are nodes on this graph. If two forwarders are overlapping, an edge between nodes representing the two forwarders is added to the graph. Fig.~\ref{graph} shows the overlapping graph of the multicasting tree in Fig.~\ref{fig3}.

On this overlapping graph, each edge has a weight. Denote the weight of edge $i$ ($i\in{[0,e-1]}$) connecting two overlapping forwarders (say $f'$ and $f''$) as $\omega_i$, where $e$ is the total number of edges in the graph. For obtaining a short-delay trajectory, $\omega_i$ is the Euclidean distance between $f'$ and $f''$, namely,
\begin{equation}
\omega_i=d_{f',f''}=\sqrt{(x_{f'}-x_{f''})^2+(y_{f'}-y_{f''})^2+(z_{f'}-z_{f''})^2},
\end{equation}
where $(x_{f'},y_{f'},z_{f'})$ and $(x_{f''},y_{f''},z_{f''})$ are the coordinates of $f'$ and $f''$ respectively. In Fig.~\ref{graph}, the red distance symbols are edge weights achieved by (4). Via this weighted overlapping graph, by employing existing algorithms (e.g., Dijkstra's algorithm, the A$^*$ search algorithm), $m$ searches the path that connects its original forwarder to its destination forwarders with the lowest weight value.

With the selected forwarders, $m$ starts forming a seamless trajectory. Excluding the original and destination forwarders, we refer to all other selected forwarders as $m$'s trajectory forwarders. Suppose there are $n$ trajectory forwarders with the $i$th ($i\in{[0,n-1]}$) one denoted as $TF_i$. $m$ calculates the intersections between the coverage edges of two overlapping trajectory forwarders $TF_i$ and $TF_{(i+1)}$ and between the coverage edges of $TF_n$ and $F_B$. Typically two distinct intersections will be obtained. The closer to $m$'s destination $B$, called the eligible intersection (EI), is employed to form $m$'s trajectory. In detail, $m$ employs Theorem 1 to check the seamlessness of the straight line between $m$'s origin $A$ and the first EI. If seamless, the straight line forms part of $m$'s trajectory. If not, $m$ employs the scheme in Section III.A to find the intersection $T$ between the edge of $TF_1$'s coverage and the straight line connecting $A$ and $TF_1$. The trajectory $(A\rightarrow{T}\rightarrow{}$the first EI) becomes a part of $m$'s trajectory. Hereafter, for the remaining parts of $m$'s trajectory, straight lines connecting consecutive EIs are used. This is because two consecutive EIs are covered by the same trajectory forwarder, ensuring that the straight line between them is seamless. The last part of $m$'s trajectory is formed by the seamless straight line between the EI and $B$ as both locations are covered by $F_B$.

We use an example in Fig.~\ref{fig3} to show how to form such a trajectory. Based on the overlapping graph (Fig.~\ref{graph}), suppose drone 6 selects drones 3, 1, $\&$ 5 to support its transition. Drone 6 calculates the EI (denoted as $D$ in the figure) between the coverage edges of drones 1 $\&$ 5. By Theorem 1, drone 6 decides that $A\rightarrow{D}$ is seamless and hence includes it as part of the trajectory. Now, as drone 1 overlaps with the destination forwarder drone 5, ${D}\rightarrow{B}$ becomes the remaining part of the trajectory. The red dotted arrow lines show the trajectory.  

\subsection{The ETTA Algorithm}
During establishing the multicasting tree, the selected LCRT forwarders exchange location information\footnote{The location information may be obtainable for example via a GPS receiver. Research studies (e.g., [21]) also proposed good schemes to locate nodes in mobile ad-hoc networks.}, and  calculates and exchanges their Euclidean distance to each other. Then, combining drone transition schemes between overlapping or non-overlapping forwarders, we present the ETTA algorithm.
\begin{tabbing}
\renewcommand{\baselinestretch}{1}
---------------------------------------------------------------------------\\[-6pt]
{\bf\small Algorithm 1 Efficient Transition via Trajectory Adjustment}
\\ xxxxxx\=xxx\=xxx\=xxx\=xxx\=xxx\=xxx\=xxx\= \kill \small Input:
\> Mobile drone $m$, $m$'s origin ($A$) and destination ($B$), \\ $m$'s origin and destination forwarders $F_A$ and $F_B$;
\\
Output: \> $m$'s ETTA transition trajectory from $A$ to $B$. \\
---------------------------------------------------------------------------\\[-6pt]
1. $m$ checks whether $F_A$ and $F_B$ are overlapping or not; \\
2. \> If overlapping, by Theorem 1, $m$ checks whether the \\ straight-line trajectory is seamless or not; \\
3. \>\> If so, $m$ transits via $A\rightarrow{B}$ directly; Exit. \\
4. \>\> If not, $m$ decides $T$ to form a new seamless traje- \\ ctory $A\rightarrow{T}\rightarrow{B}$ to transit; Exit. \\
5. \> If non-overlapping, \\
6. \>\> $m$ generates an overlapping graph for LCRT for- \\ warders; $m$ assigns weights to edges on the graph by (4); \\
7. \>\> $m$ employs Dijkstra's or A$^{*}$ algorithm to find a \\ path with the minimum weight value; suppose $n$ trajectory \\ forwarders on the path; \\
8. \>\> $m$ calculates the EI between the edges of the \\ coverage of $TF_1$ and $TF_2$; \\
9. \>\> If the trajectory ($A\rightarrow{}$this current EI) is seamless \\ based on Theorem 1, it becomes part of $m$'s trajectory; \\
10. \>\> Otherwise, $m$ calculates $T$ to form a new seam- \\ less part of its trajectory, ($A\rightarrow{T}\rightarrow{}$this current EI); \\
11. \>\> $i=2$; \\
12. \>\> While $i<n$ \\
13. \>\>\> $m$ calculates the EI between the edges of the \\ coverage of $TF_i$ and $TF_{(i+1)}$; the straight line between the \\ last EI and this EI becomes part of $m$'s trajectory; $i=i+1$; \\
14. \>\> $m$ calculates the EI between the coverage edges of \\ $TF_n$ and $F_B$; the straight line from the EI to $B$ is the last \\ part of $m$'s trajectory; Exit. \\
---------------------------------------------------------------------------\\[-6pt]
\end{tabbing}

\section{Simulation Evaluations}
\begin{table}[h]
\begin{center}
\caption{Simulation Parameters} \label{parameters1}\vspace{1em}
\begin{tabular}{|l|l||l|l|}
\hline {\bf Parameters} & {\bf Values} & {\bf Parameters} & {\bf Values} \\
\hline Frequency & 2.4GHz & Propagation model & Free space \\
\hline Dimensions & 3D & Transmission power & 15dBm \\
\hline Number of & 1 & Wireless channel & 54Mbps \\
channels &  & data rate & \\
\hline Receive & -80dBm & MAC protocol & 802.11 \\
threshold & & & \\
\hline Antenna & Omnidirectional & Simulation time & 200s \\
& antenna & & \\
\hline
\end{tabular}
\end{center}
\end{table}
We conduct experimental studies with the discrete event network simulator NS2.35 [18] to compare four multicasting schemes when they handle mobile group members: {\it LCRT} [14-15] that does not provide transition support for mobile group members; {\it T-LCRT} that enhances LCRT by selecting drones on the multicasting tree to support mobile transition. A drone receiver may forward data to a mobile drone when it is close to this receiver; {\it EGMP} [17], a geographic multicasting, that groups drones into zones and connects these zones via a bi-directional tree; {\it ETTA}, i.e., our algorithm that supports drone transitions by only using forwarders on the multicasting tree. Table I lists common settings used in our simulations. We evaluate the following performance for the four schemes. Our results plotted are the mean values of 20 simulation runs.
\begin{itemize}
\item Average multicast delay (AMD). $AMD=\frac{AD_i}{n}, i\in{[0,n-1]},$
where $AD_i$ is the average packet delay at the $i$th drone receiver, and $n$ is the total number of drone receivers in the group.
\item Average multicast throughput (AMT). $AMT=\frac{AT_i}{n}, i\in{[0,n-1]},
$
where $AT_i$ is the average data throughput at the $i$th drone receiver.
\end{itemize}

\subsection{Evaluation of Small-Group Mobile Multicasting}
\begin{figure}[h]
\begin{center}
\begin{tabular}{c}
\includegraphics[trim=40 180 40 220, clip,height=2.1in]{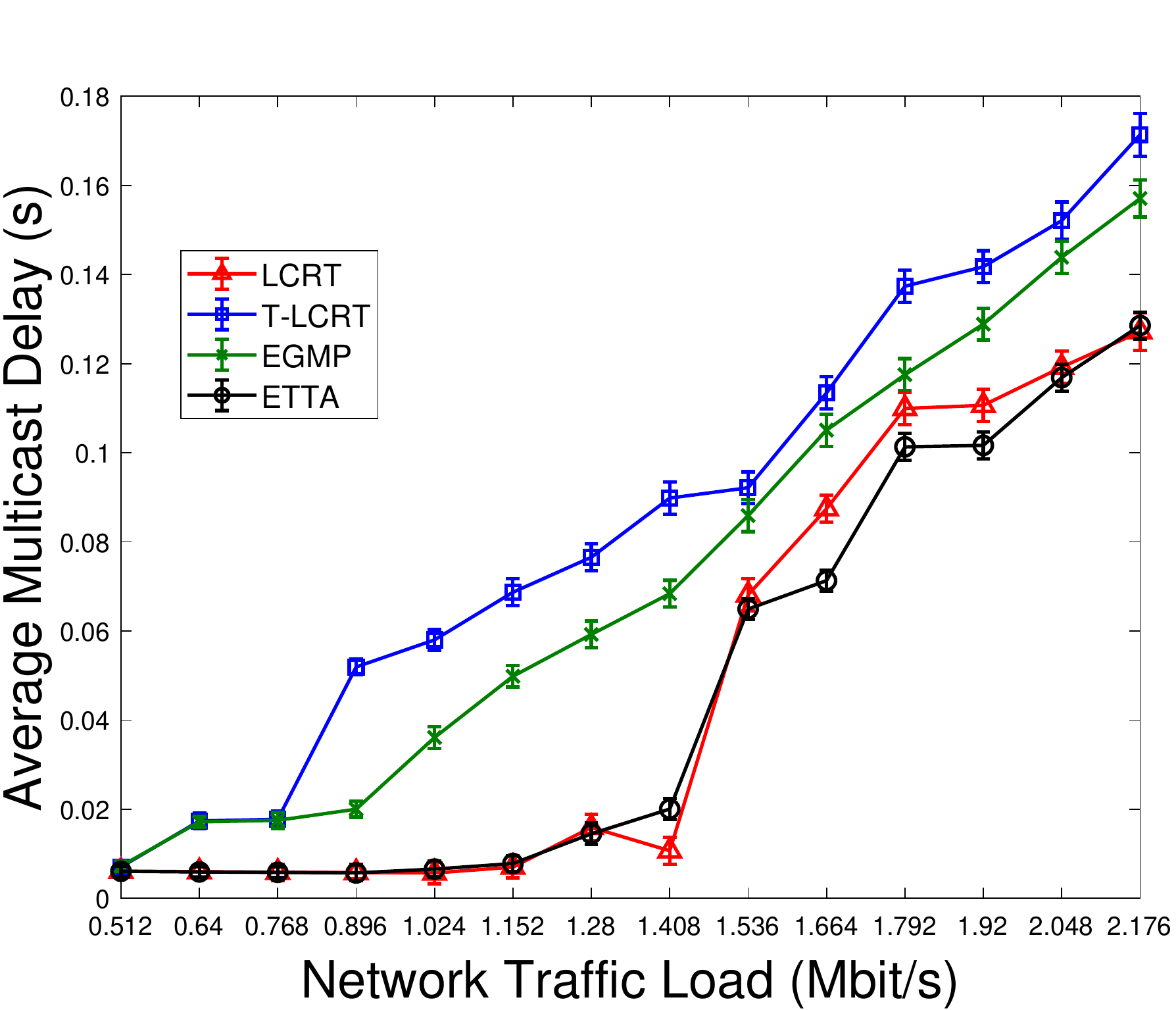}
\end{tabular}
\end{center}
\caption{Comparison of AMDs in the small-group simulation.}\label{smallgroupdelay}
\end{figure}

We first conduct a small-group simulation with 9 drones. There is 1 mobile drone which moves a distance of 102.6 meters at a speed of 10m/s during the multicasting. We vary the network traffic load from 512Kbit/s to 2.176Mbit/s to observe the four schemes. Fig.~\ref{smallgroupdelay} shows the AMD performance. LCRT and ETTA achieve shorter AMDs than T-LCRT and EGMP do. This is because T-LCRT and EGMP employ nodes that are not forwarders on the multicasting structure as transition forwarders for the mobile drone, while ETTA makes use of multicasting forwarders to handover the mobile drone and LCRT does not implement any handover process. The employment of transition forwarders that are not on the multicasting structure generates extra traffic to the system, prolonging the multicasting delays of T-LCRT and EGMP. Between T-LCRT and EGMP, T-LCRT issues control traffic to the system in order to determine suitable transition forwarders. This extra traffic load worsens AMD for T-LCRT as compared to EGMP. Both LCRT and ETTA achieve all AMDs under 150ms in this simulation. The slight AMD difference is due to the fact that they calculate AMDs based on different packets: the AMD of LCRT does not take those packets dropped during mobile transition into account while ETTA calculates the AMDs based on all transmitted packets.

\begin{figure}[h]
\begin{center}
\begin{tabular}{c}
\includegraphics[trim=40 180 40 240, clip,height=1.95in]{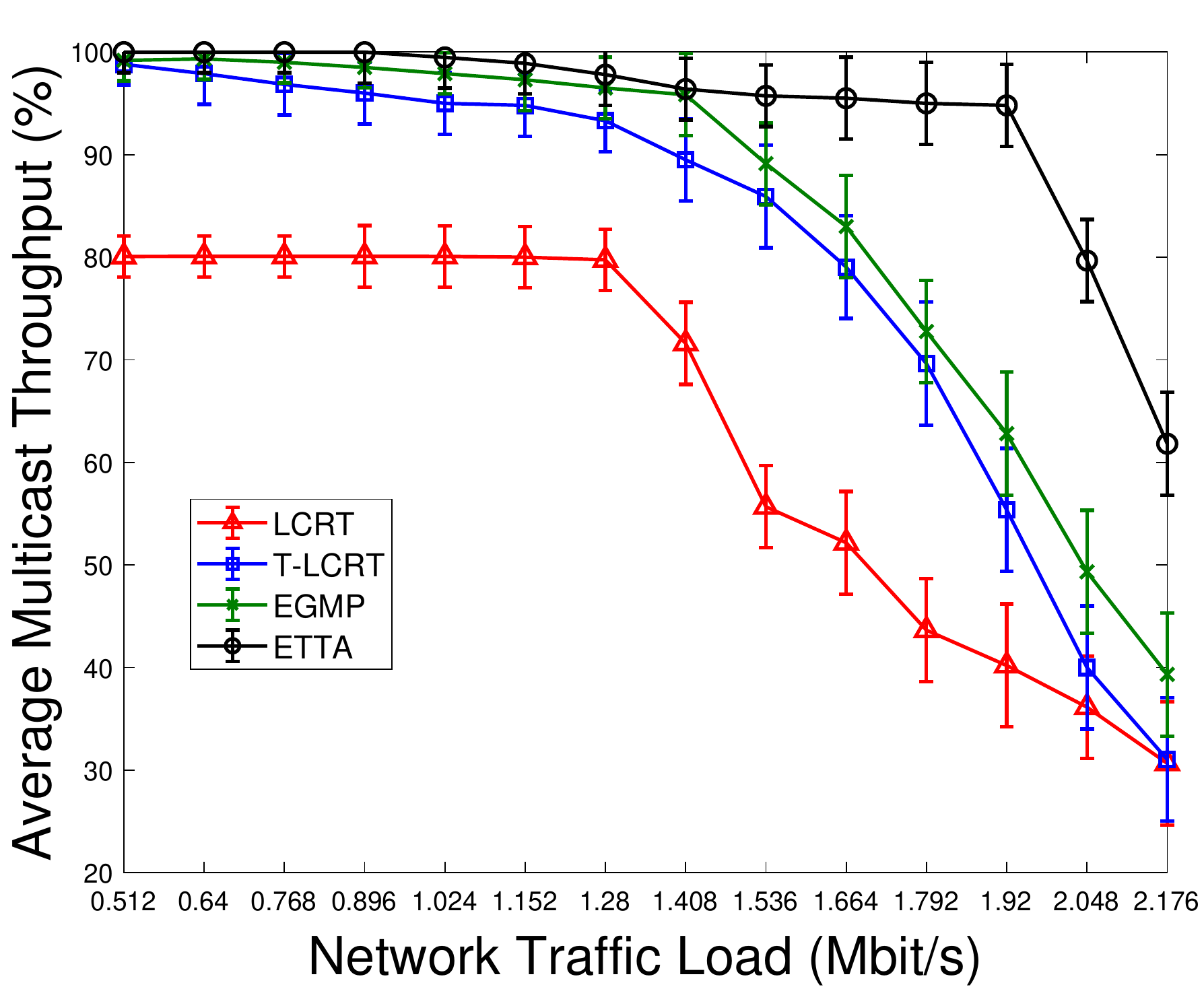}
\end{tabular}
\end{center}
\caption{Comparison of AMTs in the small-group simulation.}\label{smallgroupthroughput}
\end{figure}

Fig.~\ref{smallgroupthroughput} plots the AMT performance. ETTA achieves the highest AMT because it attempts to transit the mobile drone by planning trajectories fully covered by the LCRT tree. ETTA trajectories are formed by multicasting forwarders and hence no extra traffic is generated to the system. Also, such trajectories are planned using LCRT forwarders' coordinates that are obtained when establishing the LCRT tree, generating little control traffic to the system. For LCRT, it has the lowest AMT because the mobile drone does not receive data during its movement. EGMP and T-LCRT both employ transition forwarders to provide connections to the mobile drone, 
allowing them to achieve better AMTs than LCRT. Moreover, EGMP uses transition forwarders without changing the multicast structure and these transition forwarders can transmit to the mobile drone in time, contributing to EGMP's higher AMT than T-LCRT. This improvement is achieved by asking the mobile drone to travel around 20 meters further. 


\subsection{Evaluation of Large-Group Mobile Multicast}
\begin{figure}[h]
\begin{center}
\begin{tabular}{c}
\includegraphics[trim=40 180 40 240, clip,height=2.1in]{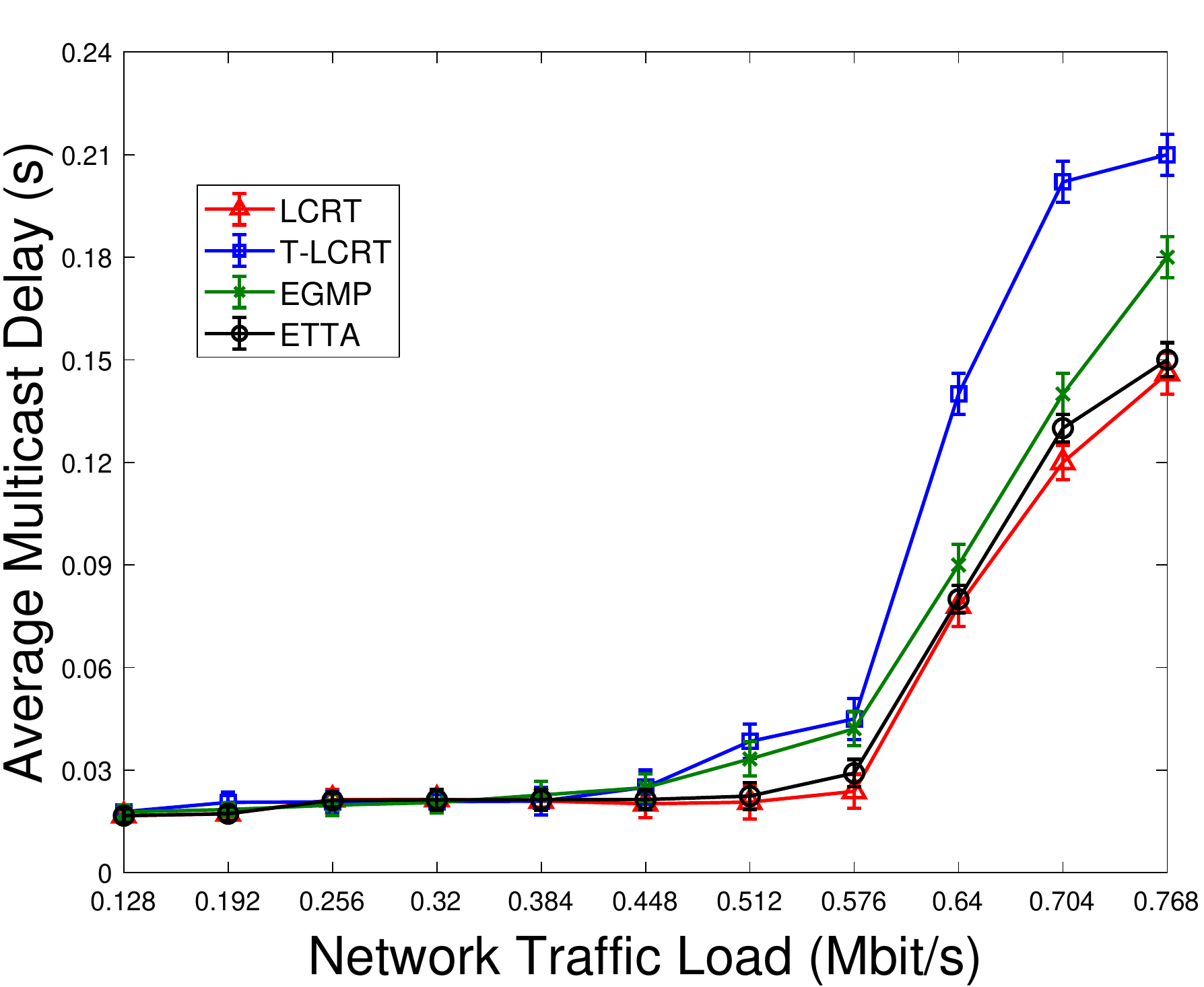}
\end{tabular}
\end{center}
\caption{Comparison of AMDs in the large-group simulation.}\label{largegroupdelay}
\end{figure}
\begin{figure}[h]
\begin{center}
\begin{tabular}{c}
\includegraphics[trim=40 180 40 240, clip,height=2.1in]{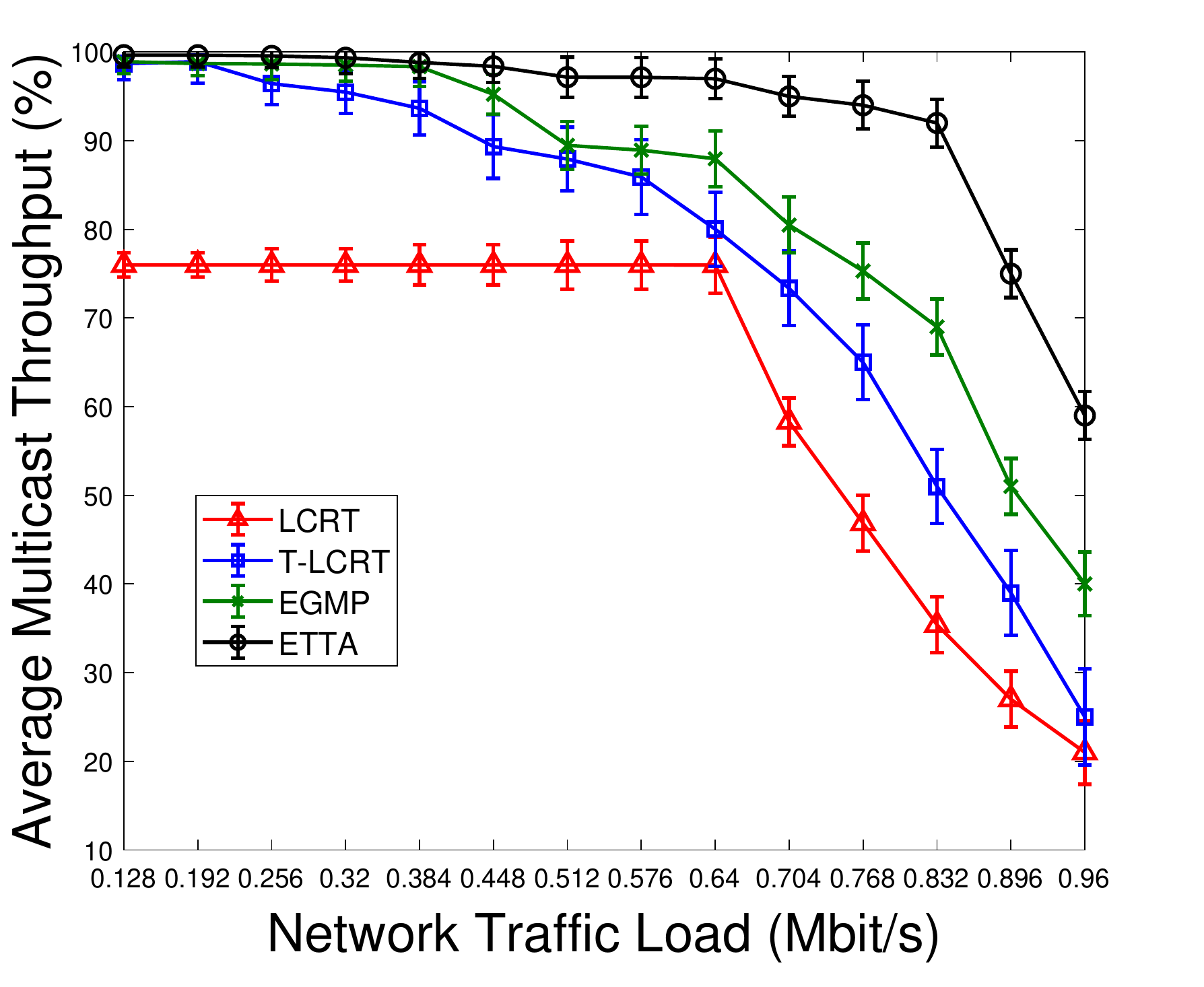}
\end{tabular}
\end{center}
\caption{Comparison of AMTs in the large-group simulation.}\label{largegroupthroughput}
\end{figure}
The large-group simulation has the 165 drones distributed so that each transmission range has 10 drones. Three mobile drones exist: the first one moves 587 meters at a speed of 10m/s, the second one moves 346 meters at a speed of 25m/s, and the third one moves 608 meters at a speed of 20m/s. We evaluate the four multicasting algorithms when the network traffic load varies from 128Kbit/s to 960Kbit/s. Based on Fig.~\ref{largegroupdelay}, the four protocols yield similar relative results for AMD in the large-group simulation as was observed in the small-group simulation. Similar reasons for the results in Fig.~\ref{smallgroupdelay} can explain the results plotted in Fig.~\ref{largegroupdelay}.

In Fig.~\ref{largegroupthroughput}, ETTA achieves higher AMT performance than other protocols. As compared to the AMT from the small-group simulation (in Fig.~\ref{smallgroupthroughput}), although the relative results are similar, ETTA outperforms other protocols by a wider margin in the large-group simulation. ETTA achieves good AMT ($90\%$ or above) when the network traffic load is $\leq$840Mb/s, while EGMP and T-LCRT achieve good AMT when the network traffic load is less than 512Mb/s and 448Mb/s respectively. In another words, ETTA carries $64\%$ or $87\%$ more traffic with guaranteed AMT than EGMP and T-LCRT. This is because in the large-group simulations, EGMP and T-LCRT require more complicated procedures or take more time to find transition forwarders. More transition forwarders issue extra data traffic to the system as well. 
This improvement is achieved by asking a mobile drone to travel around 87 meters further on average. 

\section{Conclusion}
In this paper, we studied drone multicasting in order to enable high-performance group communications between drones. Our development focused on how to seamlessly transit mobile drones in a resource-efficient manner, given the resource limitations experienced by drones and their wireless connections. A new algorithm, ETTA, was proposed that takes advantage of the obstacle-free aerial communication environment to establish straight-line trajectories for mobile drones. As such straight-line trajectories may not always be seamless, we theoretically presented the trajectory adjustment condition by which the ETTA algorithm can determine the seamlessness of a straight-line trajectory. To replace an interrupted straight-line trajectory, we proposed new schemes to form  a distance-controlled trajectory with forwarders already on the multicasting tree. As such, the ETTA algorithm allows fast drone transitions while controlling traffic overheads issued to the network. Our simulation results proved that ETTA delivers multicast data with acceptable performance when the multicasting system carries $64\%$ more traffic than compared mobile multicasting protocols.
\bibliographystyle{plain}

\end{document}